\documentclass[epjST]{svjour}
\usepackage{graphics,graphicx}
\usepackage{amsmath,amssymb,amsfonts}
\begin{document}
\title{Multivariable Optimization: Quantum Annealing \& Computation}

\author{Sudip Mukherjee\thanks{\email{sudip.mukherjee@saha.ac.in}} \and Bikas K Chakrabarti\thanks{\email{bikask.chakrabarti@saha.ac.in}}}

\institute{Condensed Matter Physics Division\\Saha Institute of Nuclear Physics\\1/AF Bidhannagar, Kolkata 700064, India }

\abstract{
Recent developments in quantum annealing techniques have been indicating potential advantage of quantum annealing for solving NP-hard optimization problems. In this article we briefly indicate and discuss the beneficial features of quantum annealing techniques and compare them with those of simulated annealing techniques. We then briefly discuss the quantum annealing studies of some model spin glass  and kinetically constrained  systems.
}

\maketitle 
\section{Introduction}
\label{sudip-intro}

           Optimization of multivariable cost functions has remained a difficult problem to solve. Situation becomes even worst when we are dealing with the NP-hard type of problem (search time is not bounded by any polynomial in total number of variables of the system). An effective proposal was forwarded by Kirkpatrick et al. \cite{sudip-sa} to tackle such optimization problems. The technique is often called the simulated annealing (SA) technique. In this technique one looks for an effective (classical) Hamiltonian, the ground state of which contains the solution. Inspired by the annealing techniques in metallurgy, one starts with the melt or high temperature (large thermal fluctuation driven) phase. Due to the presence of large thermal noise, the entire (free) energy landscape is accessible to the system (if the energy barriers are of finite height). The temperature of the system is then brought down  very slowly (annealing) and  eventually brought down to zero. During the course of annealing, the system visits 
various local (free) energy minima and comes out from them  with help of thermal fluctuations. At the end of the annealing schedule, the system is expected to be in the global minima of (free) energy, which corresponds to  global minima  of the internal energy, associated with the optimized value of the  cost function.

 The technique fails to work well in  cases where the heights of the barriers, separating local and global minima are macroscopic (of order of the system size). Search of ground state of an Ising spin glass \cite{sudip-binder} having $N$ spins (having $2^N$ states to search from) or of the minimum travel path connecting $N$ cities (Travelling Salesman Problem \cite{sudip-sa} having O($N!$) paths) on a plane are the examples of such cases: search times grow exponentially with $N$. Effective barriers here, which are classically improbable to overcome by making thermal jumps at any finite temperature, can be penetrated by quantum tunneling or fluctuations if the barriers are thin.  This idea was first suggested by Ray et al. \cite{sudip-ray} in 1989 (see Thirumalal et al. \cite{sudip-thi}, suggesting almost simultaneously similar possibilities; see also \cite{sudip-p.sen}). Hence instead of, or in conjunction with,  thermal fluctuation, if one adds quantum fluctuation in the system, such frustrated systems can regain ergodicity. Of course  
for annealing down to the respective ground state, one needs to reduce gradually the quantum fluctuation and to let it vanish eventually. The scheme is called quantum annealing (QA) \cite{sudip-finnila,sudip-kado,sudip-brooke,sudip-farhi,sudip-santoro-02,sudip-revmod}.

 Though the idea seems to be simple enough, it had not been a very straightforward issue and has been the focus of major criticisms (see e.g., \cite{sudip-boris,sudip-petter}). The basic point behind the criticisms had been that the incoherent mixture of the tunnelling waves coming from different energy barriers,  may lead to localization of the wave functions while penetrating energy barriers, as the phases of the transmitted waves are completely random. However several recent theoretical and experimental studies (see e.g., \cite{sudip-farhi,sudip-santoro-02,sudip-revmod,sudip-das,sudip-santoro,sudip-morita,sudip-torres,sudip-dwave,sudip-nasa,sudip-standford,sudip-harvard,sudip-bkc-book,sudip-bose,sudip-asim,sudip-boxio,sudip-helmut,sudip-cohen,sudip-rajak,sudip-davide,sudip-ortiz,sudip-eleanor,sudip-Dutta-book}) seem to have cleared almost all such doubts. Recently the technological implementation of QA to develop an analog quantum computer has been mastered successfully  by the D-wave systems \cite{sudip-dwave}. The performance of such computing machines, often called quantum annealers, seem to have passed the tests of quantum nature in the computation, though not very 
satisfactory yet in the context of speed enhancement (compared to the classical computers) so far for generalized cases of computational problems (see e.g., \cite{sudip-boxio,sudip-helmut}).

\section{Advantage due to Quantum tunneling in QA}
\label{sudip-aqa}
      In QA scheme, the cost function of a multivariables of optimization problem is mapped on to the energy function corresponding to a classical (frustrated) Hamiltonian ($H_0$). A time dependent quantum kinetic term ($\lambda (t)H'$)  is then added to the system. When $H'$ is non-commuting with the $H_0$, it provides quantum dynamics to the system. The total time dependent Hamiltonian becomes $H (t) = H_0+\lambda (t)H'$. The evolution of the system is visualized by solving the time dependent Schr$\ddot{\text o}$dinger  equation of the system
\begin{align*}
i{\hbar}{\frac{\partial \psi}{\partial t}} = \Big[ \lambda (t)H'+H_0\Big]\psi .
\end{align*}
If $\lambda (0)$ is taken to be very large, then $\psi$ starts effectively as the ground state of $H'$, which is assumed to be known. As $\lambda (t)$ starts decreasing slowly enough then, following the quantum adiabatic theorem, the system will be carried into the ground state of the instantaneous total Hamiltonian. At the end of the annealing schedule the kinetic term becomes zero ($\lambda (t) = 0$). Hence, one would expect the system will arrive at the ground state of $H_0$, thereby giving the optimized value of the original cost function.

\begin{figure}[h]
\begin{center}
\includegraphics[width = 7cm]{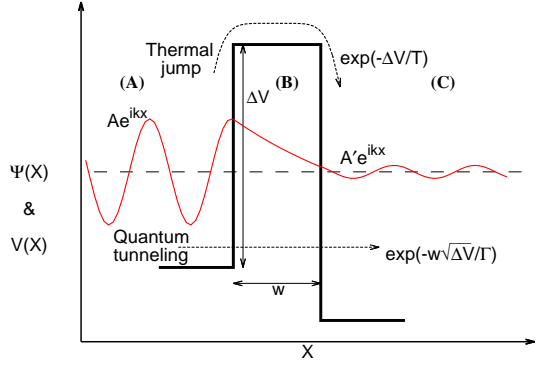}
\caption{Schematic diagram indicating the advantage of quantum tunneling over thermal hopping across a potential barrier. For kinetic energy ($E$) greater than the potential ($V$), as in regions (A) and (C), the plane wave functions ($\psi (X)  =  A e^{ikX}$ or $A' e^{ik'X}$) are indicated with wave vectors $k = \sqrt{E-V} \approx k'$ in the two regions respectively (assuming $V \approx V'$ for them). For region (B), the wave vector becomes imaginary (as $\Delta V > E$) and the damped amplitude $A'$ in (C) region will be given by $A e^{-w\sqrt{\Delta V - E}}$.} 
\label{sudip-barrier}
\end{center}
\end{figure}

The phenomenon of quantum tunneling  suggests a  better performance of QA with respect to SA in cases of high but narrow  energy barriers. Such intuitive expectation can be made a little quanitative using the following argument: if the system tries to overcome an energy barrier of height ${\Delta}v$, the classical probability of escape over the barrier is of the order of $\exp{ (-\frac{\Delta v}{T})}$, where $T$ denotes the temperature of the system, while  the quantum tunneling probabity is of the order of $\exp{ (-\frac{w\sqrt{\Delta v}}{\Gamma})}$, where \textquoteleft$w$\textquoteright ~denotes the width (assumed to be independent of barrier height $\Delta v$) of the barrier and $\Gamma$ denotes the kinetic energy  or the strength of the quantum fluctuation. One can thus realize the advantage of QA over SA due the factor $\sqrt{\Delta v}$ (instead of $\Delta v$) in the exponent of the escape probability. More explicitly, for overcoming a barrier of $N$ spins interacting with the normalized strength and each having kinetic energy $\Gamma$, the barrier height $\Delta V$ scales as $N$ while the kinetic energy $E$ scales as $N\Gamma$. Hence with wave vectors $k \approx k'  =  \sqrt{E- V}$ (in the regions A \& C in Fig. \ref{sudip-barrier}), the barrier transmission probability,  given by the damping factor $(\frac{A'}{A})^2$ in  Fig. \ref{sudip-barrier}, scales as $\exp{(-w\sqrt{N(1-\Gamma)})} \sim \exp{ (-\frac{\sqrt{N}}{\Gamma})}$ as  $w$ becomes independent of $N$ (while for classical case the probability scales as $\exp(-{\frac{N}{T}})$). With annealing schedules $T(t) = 1-\frac{t}{\tau_C}$ for SA ($t < \tau_C$) or $\Gamma (t) = 1-\frac{t}{\tau_Q}$ for QA ($t < \tau_Q$) one gets the integrated escape or transmission probabilities of the order of unity within the annealing time $\tau$, if $\tau_C e^{-N}\int_{0}^{1}e^{-Nx}dx = 1$ for SA or $\tau_Q e^{-\sqrt{N}}\int_{0}^{1}e^{-\sqrt{N}x}dx = 1$ for QA. This gives $\tau_C \sim e^{N}$ for SA and $\tau_Q \sim e^{\sqrt{N}}$ for QA. This $\sqrt{N}$ order advantage in quantum search, over a classical $N$  order search in a typical catalogue search type problem, was first demonstrated using Grover algorithm \cite{sudip-grover} (see also \cite{sudip-roland}).

\section{Some examples of QA}
\label{sudip-examples}
 We discusss two applications of QA (see e.g., \cite{sudip-rajak,sudip-das} for details). First we review the performance of QA annealing for the Sherrington-Kirkpatrick (SK) spin glass model. The classical SK model \cite{sudip-binder} is known to contain highly rugged energy landscape and QA seems to work here satisfactory \cite{sudip-rajak}. Next we focus on the QA of kinetically constrained system (KCS). Due to the presence of multiple constrains, this kind of system exhibits complex relaxation behaviour and essentially SA process becomes very slow. In KCS, we will discuss (following \cite{sudip-das}) how QA performs better than SA.

\subsection{QA of spin glass}
\label{sudip-spin-glass} 
Spin glasses are random magnetic models, where the spin-spin interactions are random and competing. The  signs of such interactions for different pairs of spins are not identical and even in some cases the strength of the interactions are also different. We pay our attention mostly on Sherrington-Kirkpatrick (SK) model \cite{sudip-binder}, where  the interactions are long ranged and distributed according to a Gaussian distribution. There is another well-known model, which is called Edward-Anderson (EA) model \cite{sudip-binder}. In this model again interactions follow Gausssian distribution but they are short ranged. Due to the presence of competing interactions, the spatial average of single site magnetization is zero. But above some crtical temperature (spin glass transition temperature ($T_C$)) the temporal average of single site magnetization also vanishes. Below that temperature $T_C$, i.e., in the spin glass phase, the (free) energy landscape is highly uneven. Several local (free) energy minima are seperated by 
very high (free) energy barriers.  The system gets trapped in any of those local minima and stays in it for a long time. For different realizations of interactions, system settles itself into different local minima. Hence system can explore a very small portion of its configurational space (becomes nonergodic in the thermodynamic limit).     

To perform QA on SK spin glass, a time dependent transverse field is introduced to the system. This transverse field  can flip   spins and allow the system to go from one configuration to another configuration. For large values of the transverse field this certainly happens and at the end of annealing, the system is expected to be in its global minima with high probability. Kadowaki and Nishimori \cite{sudip-kado} first reported the success of QA in SK spin glass. Shortly after their publication Brooke et al. \cite{sudip-brooke} experimentally showed the advantage of QA in spin glass systems.  We now discuss briefly a very recent work of  Rajak and  Chkrabarti \cite{sudip-rajak} on QA of SK spin glass.

In their paper Kadowaki and Nishimori \cite{sudip-kado} studied the QA of SK spin glass in presence of small constant longitudinal field. They used such small longitudinal field to destroy the trivial degnerecies of ground and excited states of SK spin glass system. These degnerecies arise due to the up-down symmetry of the spin system. But canonically, keeping such symmetry breaking field through out the entire annealing process destroys the key frustration effects and induces some error in the final result. Rajak and  Chkrabarti \cite{sudip-rajak} tuned the longitudinal field along with the transverse field in a similar fashion in their annealing. They  made a numerical study of SK spin glass with time dependent transverse and longitudinal field. They  worked again with very small system: $N = 8$ number of spins. The total time dependent Hamiltonian of their model is\\
\begin{align}
H (t) = -\sum_{\langle i,j\rangle} J_{ij}\sigma_i^z\sigma_j^z-{\Gamma} (t)\sum_{i = 1}^N\sigma_i^x-h (t)\sum_{i = 1}^N\sigma_i^z ,   
\end{align}
where $\sigma_i^z$, $\sigma_i^x$ are the $z$ and $x$ components of Pauli matrices respectively. $J_{ij}$s denote the random spin-spin interactions and they are distributed following the  Gaussian distribution $\rho (J_{ij}) = \Big (\frac{N}{2{\pi}J^2}\Big)^{\frac{1}{2}}\exp\Big (\frac{-NJ_{ij}^2}{2J}\Big)$. $\Gamma (t)$ and $h (t)$ denote the transverse and longitudinal field respectively. They have considered a very high initial values of $\Gamma (t)$ and $h (t)$ and finally made both of them equal to be zero by tuning them slowly. They had chosen two types of annealing schedules for $\Gamma (t)$ and $h (t)$. In one of the schedules, it is $\Gamma (t) = \frac{\Gamma_0}{\sqrt{t}}$ and $h (t) = \frac{h_0}{\sqrt{t}}$, where $\Gamma_0$, $h_0$ are the initial ($t = 1$) value of transverse and longitudinal field respectively. In the other schedule time variations are given by $\Gamma (t) = \frac{\Gamma_0}{t}$, $h (t) = \frac{h_0}{t}$. In both these types of schedules, when $t\to \infty$ both transverse and 
longitudinal fields become zero. To get the time evolution of the system, they had solved the time dependent Schr$\ddot{\text o}$dinger equation with the Hamiltonian ($1$). Solution of such Schr$\ddot{\text o}$dinger equation provides the instantaneous state $|\psi (t)\rangle$ of the system. By employ exact diagonalization technique for small system size ($N = 8$), they got the initial ground state $|\psi_0\rangle$ of the Hamiltonian ($1$) with $\Gamma (t)$  =  0 = $h (t)$. Starting from arbitrary spin state (para state) then, they evolved the system according to the Schr$\ddot{\text o}$dinger equation and calculated the instantaneous overlap of $|\psi (t)\rangle$ with $|\psi_0\rangle$, which is given by $P (t) = |\langle \psi_0|\psi (t)\rangle|$. 

\begin{figure}[h]
\begin{center}
\includegraphics[width = 12cm]{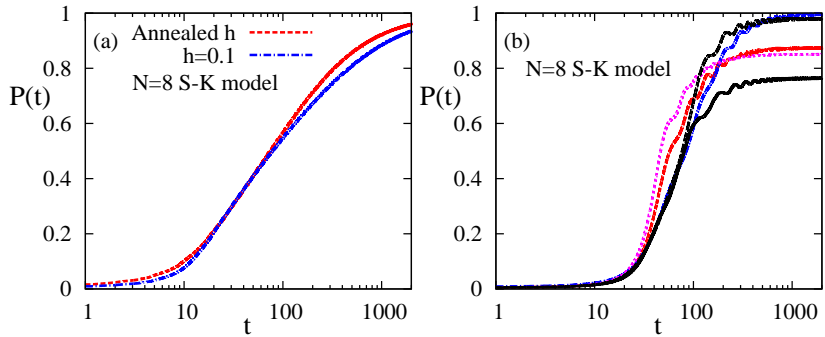}
\caption{(a) Time growth of overlap ($P(t)$) of $|\psi (t)\rangle$ with $|\psi_0\rangle$, starting from para phase at $t = 1$, with $\Gamma (t)  =  \frac{3}{\sqrt{t}}$ and $h(t)  =  -\frac{0.5}{\sqrt{t}}$ in Hamiltonian ($1$), with which the Schr\"{o}dinger equation is numerically solved. For comparison, the same results for $h(t)  =   0.1$ is given as well (which attains a bit lower value of $P(t)$ for large $t$). (b) Time growth of $P(t)$ for $5$ nominally identical realizations of exchange interactions in the same SK model with $\Gamma (t)  =  \frac{4}{t}$ and $h(t)  =  -\frac{1}{t}$ (from \cite{sudip-rajak}).} 
\label{sudip-annealing}
\end{center}
\end{figure}
 The numerical results for the time variation of $P(t)$, the overlap of $|\psi (t)\rangle$ obtained from numerical solution of the Schr$\ddot{\text o}$dinger equation with Hamiltonian ($1$) and $\Gamma (t)  =  \frac{3}{\sqrt{t}}$, $h(t)  =  -\frac{0.5}{\sqrt{t}}$, with $|\psi_0\rangle$ obtained from the exact solution of the ground state for such small system ($\Gamma  =  0  =  h$), are shown in Fig. \ref{sudip-annealing}(a). Here, a comparison with the fixed but small field ($h(t)  =  0.1$) is also shown. The advantage of the tuning of the longitudinal field along with the transverse field is clearly seen \cite{sudip-rajak}. The same advantage remains true for different spin glass realizations (see Fig. \ref{sudip-annealing}(b)).

\subsection{QA of Kinetically Constrained System}
\label{sudip-kcs}
In this subsection we will discuss QA of KCS. Study of such system is important to understand how much constraints are solely responsible for complex relaxation behaviour of a system, without accounting for any complexity which  comes due to the roughness of the (free) energy landscape. Das et al. \cite{sudip-das} made a  study of such of kinetically constrained system and  compared the performance of QA and SA on such systems. They took a one dimensional chain of $N$ non-interacting Ising spins (with a specific constraint) placed in a longitudinal field. The constraint was implemented into the system through the  East model: the $i$-th spin cannot flip if (say) the $ (i-1)$-th spin is down. Hence the domain wall cannot grow on (say) the left side unless the thermal fluctuation from the left side flips the $ (i-1)$-th spin. Hence the more number of spins turn down, system takes more time to reach its equilibrium. This makes relaxation timescale exponentially large (of the order of $e^{\frac{1}{T^{2}}}$, where $T$ 
is the temperature of the system).
     
\begin{figure}[h]
\begin{center}
\includegraphics[width = 10cm]{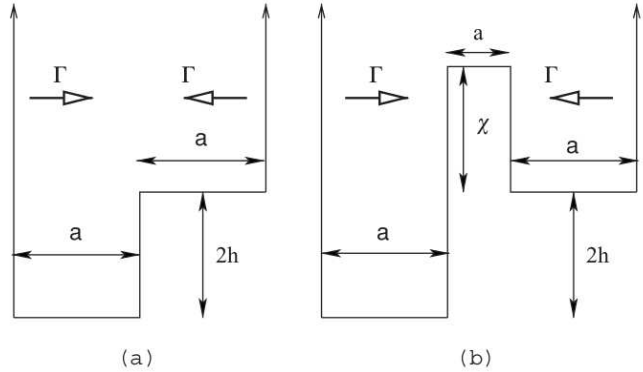}
\caption{The particle in well mapped version of the spin system in KCS : the $i$-th spin when $ (i-1)$-th is up in (a) and when the $ (i-1)$-th spin is down in (b).} \label{sudip-KCS}
\end{center}
\end{figure}

To perform QA, Das et al. \cite{sudip-das} mapped such a kinetically constrained spin system into a particle barrier crossing model. In their model each spin state is equivalently represented by a particle state in an asymmetric double-well with infinite boundary walls (particle localized in any one of the wells). If the particle is in the lower well, then the corresponding spin state is assumed to be down and vice versa. The energy difference between the two wells is $2h$ (which is exactly the energy difference between up and down state of a single spin in presence of the longitudinal field $h$). The constraint is represented by introducing  a barrier of height $\chi$ and width \textquoteleft$a$\textquoteright ~between the two wells of the $i$-th double well, when the particle of $ (i-1)$-th double-well is in the lower energy  well (see Fig. \ref{sudip-KCS}). Such a high barrier appears dynamically due to the presence of the constraint. Of course there is no such barrier when the particle of the  $ (i-1)$-th double-well is in the upper well. Any kind of quantum or thermal fluctuation may allow the particle to make transitions between the two wells of the asymmetric double-well. 

If the particle in the double-well has some quantum fluctuation (kinetic energy) $\Gamma$, then the particle will essentially scattered by the potential barrier which separates the two wells. In such situation particle has some finite probability $p$ of tunnel through the separator barrier. As the particle can be present in any one of the wells of the double-well, then the tunneling can  occur from upper well to the lower well and vice versa. They considered these probabilities by treating the situation similar to scattering of a particle by a rectangular potential barrier.  The probabilities of all possible situations for such mapped cases then:\\
 (i) If the $ (i-1)$-th spin is up and $i$-th spin is also up, then $p = 1$,\\
 (ii) If the $ (i-1)$-th spin is up and $i$-th spin is down, then  (a) $p = 0~ \text{for}~ \Gamma<2h$ (b) $p = \frac{4[\Gamma (\Gamma-2h)]]^\frac{1}{2}}{[\sqrt{\Gamma}+\sqrt{\Gamma-2h}]^2}~ \text {for}~ \Gamma \geq 2h$,\\
 (iii) If the $ (i-1)$-th spin is down and $i$-th spin is up, then $p = \frac{4[\Gamma (\Gamma+2h)]]^\frac{1}{2}}{[ (\sqrt{\Gamma}+\sqrt{\Gamma+2h})^2+g^2]}$,\\
 (iv) If the $ (i-1)$-th spin is down and $i$-th spin is down, then  (a) $p = 0~ \text{for}~ \Gamma<2h$  (b) $p = \frac{4[\Gamma (\Gamma-2h)]]^\frac{1}{2}}{[ (\sqrt{\Gamma}+\sqrt{\Gamma-2h})^2+g^2]}~ \text {for}~ \Gamma \geq 2h$.\\
Here $g = {\chi}a$. In presence of thermal fluctuation the particle has some finite Boltzmann probability ($p \approx \exp{ (-\frac{\chi}{T})}$) of crossing the potential barrier, which is placed between the two wells of the double-well.

With these probabilities they performed QA as well as SA on KCS by Monte Carlo simulations. In case of QA, they started with  high initial value of quantum fluctuation ($\Gamma_0$). Then they decreased it in an exponential schedule: $\Gamma = \Gamma_0 \exp{ (-\frac{t}{\tau_Q})}$. Here $\tau_Q$ is the time constant (which actually controls the rate of tuning of the quantum fluctuation). They took a one dimensional spin chain of $N$ non-interacting Ising spins ($\sigma_i = \pm 1$, ~~$i = 1,....N$) with periodic boundary condition. Initial state of their of simulation were completely random and eventually the initial value of the total magnetization of the system $m = \frac{1}{N}\sum_{i = 1}^N\sigma_i$ was of the oder of zero ($m_i = 0$). The parameters of their simulation were $N = 50000$, $g = 100$ and $\Gamma_0 = 100$. 

 They had found that the system was unable to reach its ground state with $\tau_Q = 2000$. When they raised the value of $\tau_Q$ upto a very high value (e.g., $\tau_Q 
\gtrsim 5000$), then the system comfortably achieved its ground state ($m_f = 1$) within about $10^5$ time steps. They were interested in finding the optimum value of $\tau_Q$ ($ (\tau_Q)_{min}$) for a given value of $g$. They set the desired degree of annealing as $m_f = 0.8$. They  clearly found that $ (\tau_Q)_{min}$ increases  with the increase of $g$.  

 Like QA, they took an exponential schedule to tune the thermal fluctuation $T$ ($T = T_0 \exp{ (-\frac{t}{\tau_C})}$, where $T_0$ is the initial value of $T$ and $\tau_C$ is the classical time constant) during the course of SA. They had explored that, with same value of the parameters (they also took identical magnitudes of the initial value of the quantum and classical fluctuation, i.e., $\Gamma_0 \approx T_0$) and same initial configuration SA took larger time to achieve same order of annealing with respect QA. To get the final order $m_f \sim 0.9$, SA required a time constant $\tau_C \sim  10^3 \tau_Q$ and to achieve $m_f \sim 1$ such relation became $\tau_C \sim  10^4 \tau_Q$. One could thus clearly realize the advantage of QA in such KCS. They also observed another benefit of QA in case of the variation of $ (\tau_Q)_{min}$ with $g$. Unlike QA, $ (\tau_C)_{min}$ had grown exponentially with $g$ (more precisely $ (\tau_C)_{min}$ increases due to increase of $\chi$ and $ (\tau_C)_{min}$ is independent of 
barrier width $a$).    

\section{Summary and Conclusion}
\label{sudip-summary}
 The difficulties in optimization of NP-hard problems are still expected to be successfully tackled by QA, rather than with SA, although by using QA techniques, we still do not find that the search time  becomes polynomial bounded in the system size ($N$). We clearly find some indications of advantage of QA with respect to SA in context of search time: Boxio et al. \cite{sudip-boxio} indicates that the search time scales as $\exp(N^x)$ ($x \lesssim 1$; the heuristic argument presented in the Sect.~\ref{sudip-aqa} suggests $x = \frac{1}{2}$ for QA), while in case of SA such search time scales as $\exp(N)$. In spite of several criticisms, QA techniques are developing rapidly with  support from both theoretical and experimental investigations.

  Following the initial indication by Ray et al. \cite{sudip-ray}, the experimental studies of QA by Brooke et al. \cite{sudip-brooke} (see also Torres et al. \cite{sudip-torres}) and successive theoretical investigations by  Finnila et al. \cite{sudip-finnila}, Kadowaki and Nishimori \cite{sudip-kado} and Farhi et al. \cite{sudip-farhi} had shown convincingly  the probable advantage of QA technique in solving NP-hard optimization problems. Detailed mathematical reviews of QA by Santoro and Tossati \cite{sudip-santoro}  and Morita and Nishimori \cite{sudip-morita} and general review of QA and its applications in the computational field, by Das and Chakrabarti \cite{sudip-revmod} had set  an encouraging scenario. Precise hardware developments implemented in the D-wave's  quantum annealer \cite{sudip-dwave,sudip-nasa,sudip-standford,sudip-harvard,sudip-boxio,sudip-helmut,sudip-cohen,sudip-davide,sudip-ortiz,sudip-eleanor} have opened a new paradigm for analog quantum computation.

\section{Acknowledgments}
\label{sudip-ack}
 We are grateful to our collaborators A. Chakrabarti, A. Das, A. Rajak, P. Ray and R. B. Stinchcombe for their contributions at different stages of these studies.

\end{document}